\documentclass[journal]{IEEEtran}
\usepackage{cite}
\usepackage{amsmath,amssymb,amsfonts}
\usepackage{graphicx}
\usepackage{textcomp}
\usepackage{xcolor}
\usepackage{booktabs}
\usepackage{multirow}
\usepackage{hyperref}

\usepackage{booktabs}
\usepackage{xcolor}
\usepackage{colortbl}
\definecolor{HeaderBlue}{RGB}{30,60,100}
\definecolor{SoftCream}{RGB}{252,250,245}
\usepackage{array}
\usepackage{makecell}
\begin{document}

\title{Beyond the RF Paradigm: Rydberg Atomic Receivers for Next-Generation IoT}

\author{Qihao Peng,
        Qu Luo,~\IEEEmembership{Member,~IEEE}, Dongnan Xia, Zhehua Zhang,  Zeyan Zhang, 
        Jizhou Wu,\\
         Kezhi Wang,~\IEEEmembership{Senior Member,~IEEE}, Cunhua Pan,~\IEEEmembership{Senior Member,~IEEE},
        Maged Elkashlan,~\IEEEmembership{Senior Member,~IEEE},  \\
         Pei Xiao,~\IEEEmembership{Senior Member,~IEEE}, Derrick Wing Kwan Ng,~\IEEEmembership{Fellow,~IEEE},
        Trung Q. Duong,~\IEEEmembership{Fellow,~IEEE},\\
        George K. Karagiannidis,~\IEEEmembership{Fellow,~IEEE},
        Jiangzhou Wang,~\IEEEmembership{Fellow,~IEEE},
\thanks{Q. Peng, Q. Luo, and P. Xiao are affiliated with 5G and 6G Innovation Centre, Institute for Communication Systems (ICS) of University of Surrey, Guildford, GU2 7XH, UK. (e-mail: \{q.peng,q.u.luo,p.xiao\}@surrey.ac.uk).}
\thanks{D. Xia, C. Pan, and J. Wang are with the National Mobile Communications Research Laboratory, Southeast University, Nanjing, China. (e-mail: \{dnxia,cpan,j.z.wang\}@seu.edu.cn).}
\thanks{Z. Zhang, Z. Zhang, and J. Wu are with State Key Laboratory of Quantum Optics Technologies and Devices, Institute of Laser Spectroscopy, Shanxi University, Taiyuan 030006, China (e-mail: \{202412603032, 202422603017\}@email.sxu.edu.cn;wujz@sxu.edu.cn).}
\thanks{K. Wang is with the Department of Computer Science, Brunel University of London, UB8 3PH London, U.K. (e-mail: Kezhi.Wang@brunel.ac.uk).}
    \thanks{M. Elkashlan is with the School of Electronic Engineering and Computer Science, Queen Mary University of London, E1 4NS London, U.K.(e-mail: maged.elkashlan@qmul.ac.uk).}
        \thanks{Derrick Wing Kwan Ng is with the School of Electrical Engineering and Telecommunications, University of New South Wales, Sydney, NSW 2052, Australia (e-mail: w.k.ng@unsw.edu.au).}
    \thanks{T. Q. Duong is with the Faculty of Engineering and Applied Science, Memorial University, St. John’s, NL A1C 5S7, Canada. (e-mail: tduong@mun.ca.)}
        \thanks{G. K. Karagiannidis is with the Department of Electrical and Computer Engineering, Aristotle University of Thessaloniki, 541 24 Thessaloniki, Greece (e-mail: geokarag@auth.gr).} 
        }


\maketitle

\begin{abstract}
Next-generation Internet-of-Things (IoT)  is evolving toward a ubiquitous, ultra-low-power, and multi-band heterogeneous networking paradigm that seamlessly integrates terrestrial, non-terrestrial, and ambient devices. This vision places unprecedented demands on conventional radio frequency (RF) receivers, whose fundamental bottlenecks in sensitivity, power consumption, coverage, and multi-band operation are rooted in the RF antenna. To tackle these issues, we show that the quantum properties of Rydberg atomic quantum receivers (RAQRs), including ultra-high sensitivity, broad frequency agility, and diverse reception modalities, provide a physically distinct receiver-side path that replaces the conventional antenna-and-low-noise-amplifier chain. Using LoRa, narrowband IoT, and ambient IoT as case studies, this article shows that RAQRs deliver significant gains in weak-uplink, low-power, and battery-free regimes. A stochastic-geometry analysis in cellular and cell-free architectures then maps these device-level gains onto network coverage, where the RAQR retains roughly a 4 dB half-coverage advantage over the RF receiver in sparse deployments at \(\lambda \sim 10^{-5}~{\mathrm m}^{-2}\), with the gain eroded as device density grows. The open challenges are presented to stand between current RAQR prototypes and deployable IoT infrastructure.
\end{abstract}


\section{Introduction}
The next-generation Internet-of-Things (IoT) is evolving into a ubiquitous and intelligent fabric that interconnects tens of billions of heterogeneous devices. This evolution is driven by converging directions in the 3GPP ecosystem, including low-power wide-area (LPWA) technologies such as narrowband IoT (NB-IoT) and enhanced machine-type communications (eMTC), non-terrestrial networks (NTN) for extended and global coverage, and ambient IoT for battery-free or near-battery-free operation. These directions collectively push IoT networks toward lower device power, wider-area deployment, and more diverse service requirements. Consequently, realizing the next frontier of IoT depends not only on massive connectivity but also on weak-uplink connectivity under extremely stringent device-side energy constraints.

Next-generation IoT networks, however, expose fundamental limitations in conventional RF-based architectures. Weak, sporadic, and power-limited uplink signals must be reliably detected from devices that are typically low-cost, energy-constrained, and deployed in harsh propagation environments. These requirements expose intrinsic physical and architectural bottlenecks, as detailed below.
\begin{itemize}
    \item \textbf{Sensitivity}: The sensitivity of classical radio-frequency (RF) receivers is fundamentally constrained by the Johnson–Nyquist thermal noise floor of approximately \(-174\)~dBm/Hz at room temperature and is further degraded by the noise figure and implementation loss of the RF front end. In compact IoT gateways and terminals, electrically small antennas are additionally subject to the Chu-limit-induced efficiency–bandwidth trade-off, which constrains the effective received signal power. Consequently, weak uplink signals from remote, deeply shadowed, or ultra-low-power IoT devices are readily buried below the receiver noise floor \cite{centenaro2016long}.
    \item \textbf{Device-side power constraints}: IoT terminals are subject to stringent size, weight, and power (SWaP) constraints, as they are typically compact, low-cost, and powered by small batteries or harvested energy \cite{saad2019vision}. These constraints limit the antenna aperture, RF front-end complexity, transmit power, and active communication time. Consequently, massive IoT connectivity is often dominated by short, low-duty-cycle, and power-limited uplink transmissions, making reliable reception of weak signals especially challenging for ambient-powered devices.
  \item   \textbf{Scalability and heterogeneity}: Next-generation IoT networks accommodate highly diverse devices, ranging from NB-IoT and eMTC to battery-less ambient devices and NTN terminals \cite{vaezi2022cellular}. These devices differ substantially in bandwidth, power budget, and access behavior, while classical RF receivers are typically optimized for specific bands, waveforms, and operating conditions, struggling to simultaneously support massive sporadic access, wide dynamic range, multi-band operation, interference management, and heterogeneous quality-of-service requirements. 
\end{itemize}

These intrinsic physical and architectural limitations of RF receivers are difficult to address through conventional remedies alone, such as increasing terminal transmit power, enlarging antenna apertures, or adopting lower-noise but more complex RF front ends, each of which faces inherent trade-offs with the SWaP and cost constraints that define IoT. Rydberg atomic quantum receivers (RAQRs) offer a complementary, receiver-side path beyond these barriers by converting field-induced atomic responses into optical readouts. Particularly, by exploiting the giant electric-field susceptibility of high-$n$ atomic states (proportional to $n^{7}$) and reading out RF information through electromagnetically induced transparency (EIT) and Autler-Townes splitting (ATS), weak electric field perturbations on Rydberg atoms can be detected using optical measurements, thereby enabling the reliable reception of signals from IoT devices. 

Motivated by these compelling properties, deploying Rydberg receivers at base stations (BSs) or onboard satellites holds promise for substantial gains in next-generation IoT systems. The remainder of this paper is organized as follows. Section II introduces the physical mechanisms of RAQRs and explains how each property addresses the bottlenecks identified above. Section III then demonstrates the advantages of RAQR through two representative IoT scenarios, i.e., LPWA and ambient IoT networks. Section IV examines the role of RAQRs in reshaping IoT across both cellular and cell-free networks. Section V outlines the open research challenges on the path toward practical deployment. Finally, Section VI concludes the paper.

\section{Quantum Phenomena Underpinning RAQR for IoT Networks}
\label{sec:quantum_phenomena}

\subsection{From RF Fields to Optical Observations}
\label{subsec:quantum_bridge}

\begin{figure*}
       \centering
    \includegraphics[width=0.9\linewidth]{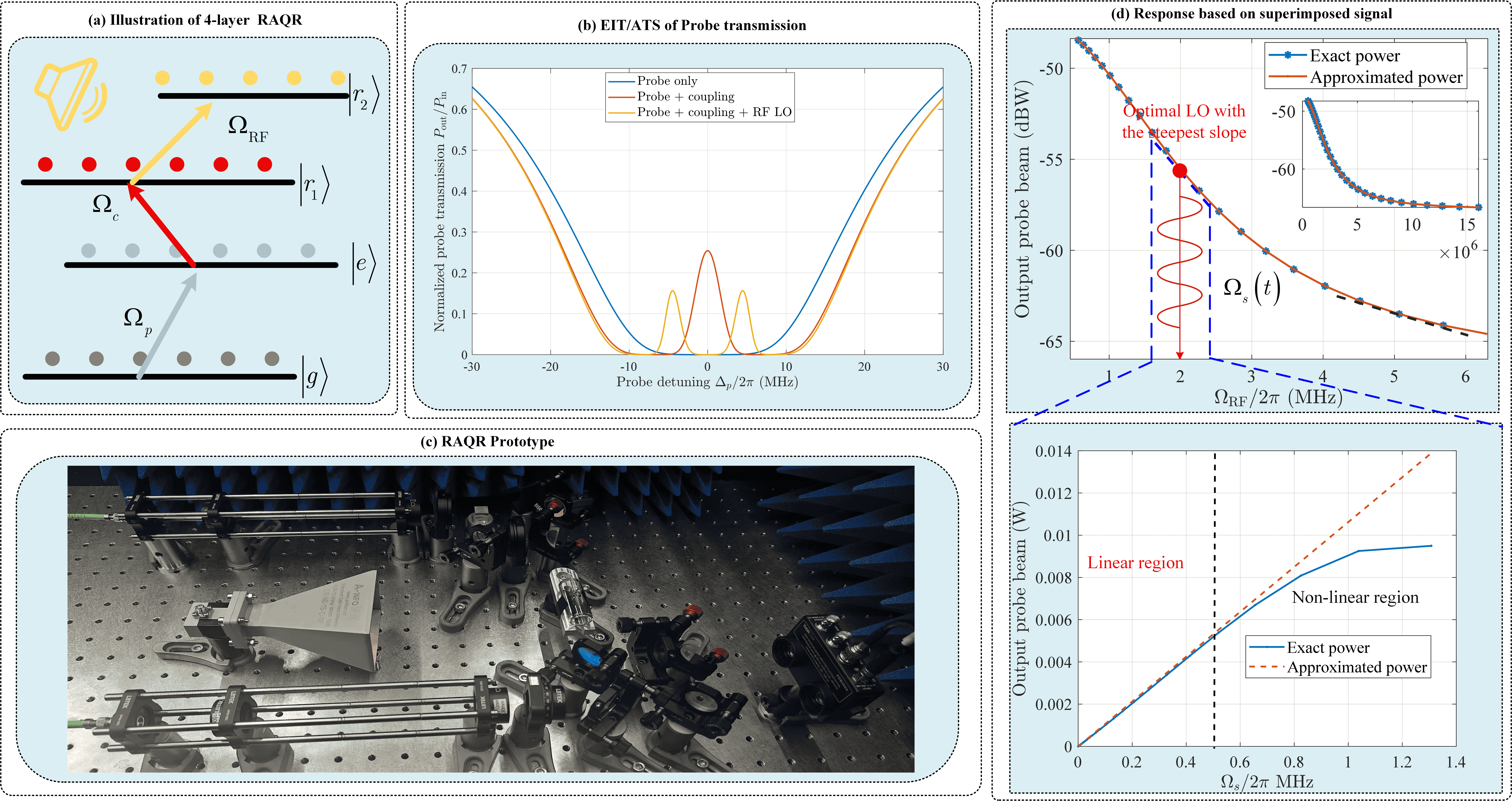}
    \caption{Principles of RAQR. (a) Four-level atomic configuration. (b) EIT and Autler-Townes-split probe transmission spectra. (c) Laboratory prototype of the RAQR setup. (d) RAQR optical response under a strong RF local oscillator, with the steepest-slope bias enabling small-signal linearization (first-order Taylor approximation).}
    \label{EIT_AT}
\end{figure*}
RAQRs convert an incident RF electric field into an optically detectable
atomic response. The conversion exploits the steep scaling of
Rydberg-state polarizability ($\propto n^{7}$) and transition dipole
moment ($\propto n^{2}$) with the principal quantum number $n$, which together render the atomic-level structure exquisitely sensitive to weak RF fields. A resonant RF field that couples two adjacent Rydberg states produces an ATS of the dressed levels, which is read out optically through the probe beam. The four-level atomic ladder configuration of RAQR is illustrated in Fig.~\ref{EIT_AT}(a), consisting of a ground state $\lvert g\rangle$, an intermediate excited state $\lvert e\rangle$, and two Rydberg states $\lvert r_1\rangle$ and $\lvert r_2\rangle$. A probe laser drives the $\lvert g\rangle \rightarrow \lvert e\rangle$ transition, a coupling laser excites $\lvert e\rangle \rightarrow \lvert r_1\rangle$, and the microwave field under detection couples $\lvert r_1\rangle \rightarrow \lvert r_2\rangle$. The microwave-induced ATS on the upper Rydberg transition imprints directly onto the probe transmission spectrum, establishing a quantum-mechanically defined mapping between the incident microwave amplitude and a measurable optical observation, as shown in Fig.~\ref{EIT_AT}(b).

Based on this mechanism, we have built a prototype of RAQR, as shown in Fig.~\ref{EIT_AT}(c). The probe beam carries the signal from the incident RF field to the optical measurement. As shown in Fig.~\ref{EIT_AT}(d), a strong RF local oscillator (LO) biases the atoms onto a steep portion of their optical response, and then the weak IoT signal acts as a small perturbation whose amplitude is read out as a proportional change in the transmitted probe power. A photodetector following the vapor cell completes the chain, mapping the optical readout back to an electrical baseband signal. The dominant noise floor is thereby shifted from the antenna chain to the atomic medium itself, where it is bounded by quantum projection and photon shot noise rather than by thermal noise. This transduction enables the RAQR to detect weak IoT signals that fall below the effective sensitivity of conventional RF receivers, establishing a quantum-defined physical layer for next-generation IoT networks.

\subsection{RAQR Opportunities for Weak-Uplink IoT}
The three bottlenecks identified call for a transformative receiver whose physical principle is decoupled from the inherent constraints of the classical RF front-ends. RAQRs overcome these bottlenecks through four complementary quantum-enabled capabilities, as summarized in Table~\ref{tab:mapping}.

\textbf{Quantum-enhanced field sensitivity}:
Leveraging the large transition dipole moments of high-\(n\) Rydberg states, RAQRs provide a receiver-side weak-field sensing mechanism that is fundamentally different from the classical antenna-based RF front end. From a fundamental perspective, it has been shown that the quantum shot-noise limit of a Rydberg RF receiver can fall below the thermal-noise floor of a conventional receiver \cite{bussey2022quantum}, especially for low-symbol-rate reception. Atomic superheterodyne reception has demonstrated \(55~{\rm nV/cm}/\sqrt{\rm Hz}\) sensitivity and a minimum detectable field of \(780~{\rm pV/cm}\) \cite{jing2020atomic}, and sensitivity can be further improved by using frequency detuning \cite{simons2016using}. More recently, a cold-atom Rydberg microwave electrometer achieved \(10~{\rm nV/cm}/\sqrt{\rm Hz}\) sensitivity only \(2.6\) times above the standard quantum limit \cite{tu2024approaching}, indicating that the RAQR is approaching its intrinsic quantum limit. Although practical RAQRs are still affected by photon shot noise, atomic projection noise, optical readout noise, thermal noise, and decoherence, experimental progress demonstrates that RAQR sensitivity is progressively approaching quantum-limited operation, making RAQR a promising infrastructure-side receiver candidate for enhancing weak-uplink reception in next-generation IoT networks.

\textbf{Infrastructure-side quantum sensing}:
RAQRs are most compelling for IoT when deployed at the infrastructure side, where the optical beams, vapor cell, and coherent detection chain are concentrated at the network node, leaving IoT terminals compact, inexpensive, and energy-constrained. This architecture brings two key benefits. First, the infrastructure-side receiver is not bound by the Chu limit, which imposes an unavoidable aperture-efficiency-bandwidth
trade-off on classical electrically small antennas, which has been experimentally demonstrated \cite{cox2018quantum}. Second, by improving the infrastructure-side ability to capture weak uplink fields, RAQRs significantly alleviate the transmit power, active duration, and repetition burden of IoT devices \cite{peng2025activebatteryfreerydbergatomic}. This is particularly relevant to LPWA and ambient IoT networks, where battery lifetime is strongly affected by coverage conditions and repeated transmissions. The result is a fundamental rebalancing of the link budget toward the infrastructure side, supporting IoT terminals that are simultaneously smaller, longer-lived, and lower-power.

\textbf{Frequency-tunable quantum reception}:
RAQRs offer a frequency-tunable reception mechanism that is well aligned with the multi-band nature of next-generation IoT networks. By choosing different alkali atoms, Rydberg states, principal quantum numbers, angular momentum states, and RF-coupled transitions, RAQRs can be configured to different carrier frequencies, ranging from direct current to THz \cite{holloway2014broadband}. Specifically, the high-angular-momentum \(nF\rightarrow nG\) Rydberg states have experimentally enabled resonant RF detection from \(800\) to \(900\)~MHz \cite{brown2023very}, directly serving the sub-GHz LPWA and ambient IoT bands. Furthermore,  it has been demonstrated that RAQRs can recover diverse modulations, including amplitude modulation, frequency modulation, and phase modulation \cite{holloway2019detecting}. Therefore, RAQRs cover the
entire IoT carrier landscape, from sub-GHz LPWA and ambient IoT
to GHz cellular bands and even NTN-IoT, requiring only the choice of atomic transition and operating point.

\textbf{Narrow bandwidth and sharp band selection}:
The narrow-band response of RAQRs, often perceived as a limitation, is in fact a structural match to next-generation IoT networks. The RF-to-optical conversion is bounded by the finite linewidth of the Rydberg-EIT response, yielding an achievable baseband bandwidth of several MHz that comfortably matches the bandwidth requirements of typical IoT services. More importantly, the narrow linewidth confers intrinsic spectral selectivity, since the atomic response becomes appreciable only when the incident RF field lies near a selected Rydberg transition, thereby physically rejecting out-of-band interference.

Together, these properties enable RAQRs to effectively address the key bottlenecks of next-generation IoT networks. Accordingly, provided that the response bandwidth, detuning, optical power, and atomic linewidth are jointly engineered, RAQRs emerge as a quantum-enabled receiver front end well aligned with, and complementary to, the conventional RF chain in next-generation IoT networks.

\begin{table*}[!t]
\centering
\caption{End-to-End Mapping: From IoT Bottlenecks to RAQR Quantum Mechanisms, Applications, and Physical Limitations}
\label{tab:mapping}
\footnotesize
\renewcommand{\arraystretch}{1.6}
\setlength{\tabcolsep}{6pt}
\arrayrulecolor{HeaderBlue}

\begin{tabular}{>{\centering\arraybackslash}p{3cm}
                >{\centering\arraybackslash}p{4cm}
                >{\centering\arraybackslash}p{3.2cm}
                >{\centering\arraybackslash}p{4cm}}
\toprule
\rowcolor{HeaderBlue}
\textcolor{white}{\textbf{IoT bottleneck}} 
& \textcolor{white}{\textbf{RAQR quantum mechanism}} 
& \textcolor{white}{\textbf{Application benefit}} 
& \textcolor{white}{\textbf{Physical limitation}} \\
\midrule

\makecell[tc]{Weak-signal reception\\ below thermal noise floor}
& \makecell[tc]{Giant electric-dipole moments;\\ Standard quantum limit}
& \makecell[tc]{NB-IoT uplink;\\ LoRa uplink;\\ Direct UE-to-satellite access}
& \makecell[tc]{Temperature drift;\\ Atomic density variation;\\ Laser-power instability} \\[2pt]

\rowcolor{SoftCream}
\makecell[tc]{Device-side SWaP\\ and energy budget}
& \makecell[tc]{Bypassing Chu-limit aperture;\\ Infrastructure-side quantum sensing}
& \makecell[tc]{Battery-free ambient IoT;\\ SWIPT-enabled terminals}
& \makecell[tc]{Photonic integration difficulty;\\ Gain replication across arrays} \\[2pt]

\makecell[tc]{Multi-band heterogeneity\\ from sub-GHz to GHz}
& \makecell[tc]{Tunable band;\\ Multi-species, multi-state tunability}
& \makecell[tc]{800~MHz LPWA;\\ 2.4/3.5/5.8~GHz ambient IoT;\\ 6.95~GHz private IoT}
& \makecell[tc]{Discrete atomic resonances;\\ Non-uniform spectral coverage} \\[2pt]

\rowcolor{SoftCream}
\makecell[tc]{Co-band interference\\ under massive access}
& \makecell[tc]{Narrow Rydberg-EIT linewidth;\\ Intrinsic spectral selectivity}
& \makecell[tc]{Out-of-band rejection;\\ SI-traceable, calibration-free}
& \makecell[tc]{Adjacent-band Stark detuning;\\ Working-point perturbation} \\

\bottomrule
\end{tabular}
\arrayrulecolor{black}
\end{table*}

\section{RAQR-enabled IoT Networks}
We quantify the system-level gains of RAQR-enabled LoRa and NB-IoT deployments and examine the potential of ambient RAQR-empowered IoT networks augmented with simultaneous wireless information and power transfer (SWIPT) under identical channel realizations and detection bandwidths.
\begin{figure*}[ht]
    \centering
    \includegraphics[width=0.9\linewidth]{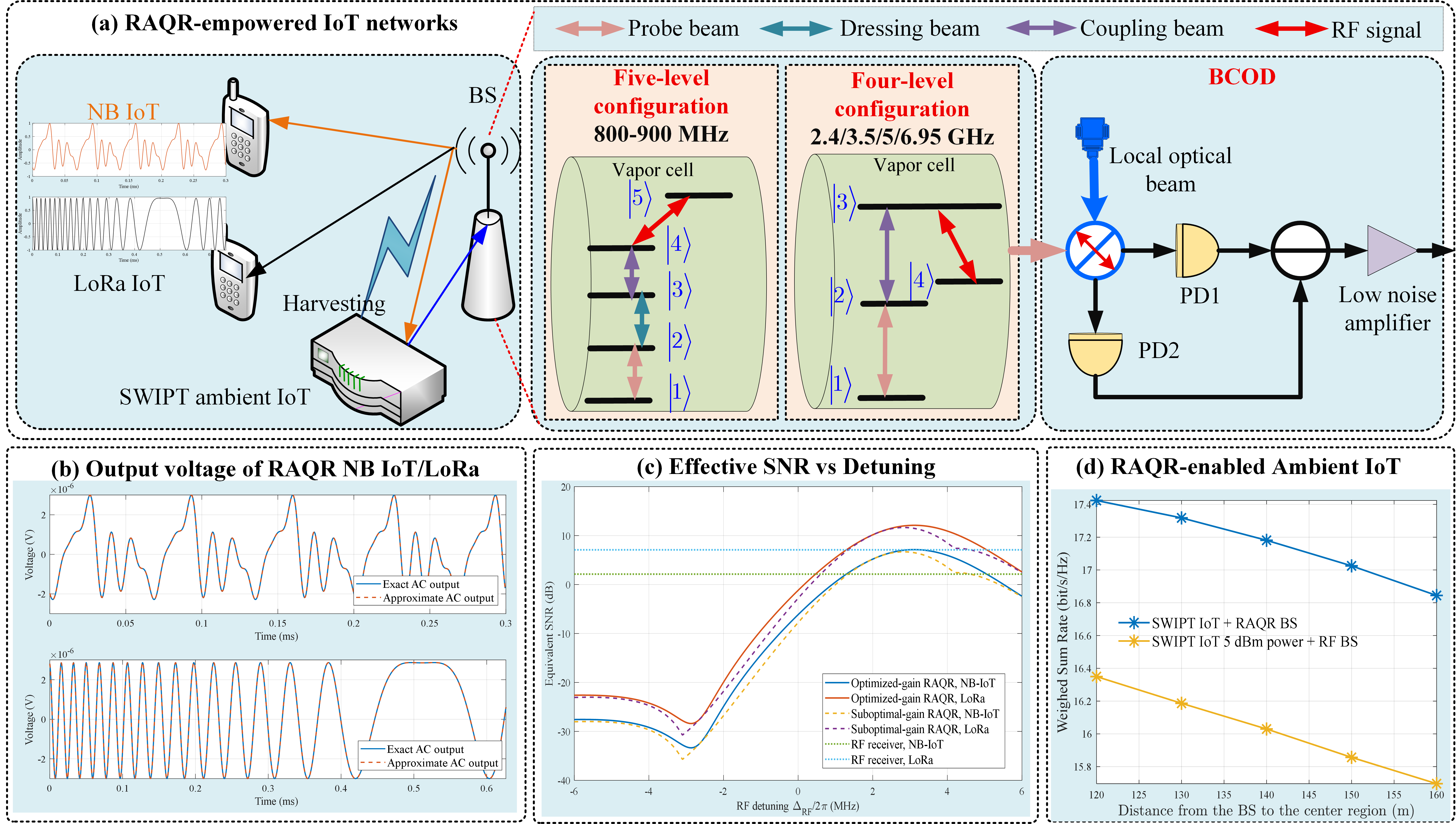}
    \caption{RAQR-enabled IoT networks. (a) Typical IoT scenario. (b) Approximated and exact waveforms. (c) Received SNR. (d) Performance comparison between RAQR and RF receiver. Simulation parameters: five-layer with \(\Omega_p = 2\pi \times 17\) MHz, \(\Omega_d = 2\pi \times 52\) MHz, \(\Omega_c = 2\pi \times 13\) MHz, \(\Omega_{LO} = 2\pi \times 9.5\) MHz, \(\Omega_{\text{IoT}} = 10^{-4} \Omega_{\text{LO}}\), 30 mW local optical beam; four-layer with \(\Omega_p= 2\pi\times 3.35\)~MHz, \(\Omega_c = 2\pi\times 0.95\)~MHz, \(\Omega_{\rm LO} = 2\pi\times  1.22\)~MHz, 30 mW local optical beam; RF baseline: NF = 5 dB on top of the thermal noise floor of (-174) dBm/Hz.}
    \label{NBIot}
\end{figure*}

\begin{table*}[ht]
\centering
\caption{Representative IoT Scenarios, Carrier/Bandwidth Settings, and Examples of Rydberg Atomic Level Configurations}
\label{tab:iot_raq_level_selection}
\scriptsize
\renewcommand{\arraystretch}{1.25}
\setlength{\tabcolsep}{4pt}

\definecolor{IEEEblue}{RGB}{0,90,156}
\definecolor{LightIEEEblue}{RGB}{232,242,250}
\definecolor{VeryLightBlue}{RGB}{245,250,253}
\definecolor{VeryLightGray}{RGB}{248,248,248}

\begin{tabular}{p{3.0cm} p{4.2cm} p{9.1cm}}
\toprule
\rowcolor{IEEEblue!18}
\textbf{IoT Scenario}
& \textbf{Carrier \& Bandwidth}
& \textbf{Atom and Representative Atomic Level Configuration} \\
\midrule

\rowcolor{VeryLightBlue}
NB-IoT, Band-20-like uplink
&
\begin{tabular}[t]{@{}l@{}}
Carrier: 832--862 MHz\\
Bandwidth: 180 kHz
\end{tabular}
&
\begin{tabular}[t]{@{}p{8.9cm}@{}}
$^{87}$Rb, five-level high-angular-momentum configuration:
$5S_{1/2}\!\rightarrow\!5P_{3/2}\!\rightarrow\!5D_{5/2}\!\rightarrow\!46F_{7/2}\!\rightarrow\!46G_{9/2}$.
The $47F_{7/2}\!\rightarrow\!47G_{9/2}$ transition can be considered for lower-800-MHz operation.
\end{tabular}
\\

NB-IoT, Band-8-like uplink
&
\begin{tabular}[t]{@{}l@{}}
Carrier: 880--915 MHz\\
Bandwidth: 180 kHz
\end{tabular}
&
\begin{tabular}[t]{@{}p{8.9cm}@{}}
$^{87}$Rb, five-level high-angular-momentum configuration:
$5S_{1/2}\!\rightarrow\!5P_{3/2}\!\rightarrow\!5D_{5/2}\!\rightarrow\!45F_{7/2}\!\rightarrow\!45G_{9/2}$.
\end{tabular}
\\

\rowcolor{VeryLightBlue}
LoRa EU868
&
\begin{tabular}[t]{@{}l@{}}
Carrier: 863--870 MHz\\
Bandwidth: 125/250 kHz
\end{tabular}
&
\begin{tabular}[t]{@{}p{8.9cm}@{}}
$^{87}$Rb, five-level high-angular-momentum configuration:
$5S_{1/2}\!\rightarrow\!5P_{3/2}\!\rightarrow\!5D_{5/2}\!\rightarrow\!45/46F_{7/2}\!\rightarrow\!45/46G_{9/2}$.
The 868-MHz carrier can be matched through RF detuning or Stark tuning.
\end{tabular}
\\

LoRa US915
&
\begin{tabular}[t]{@{}l@{}}
Carrier: 902--928 MHz\\
Bandwidth: 125/500 kHz
\end{tabular}
&
\begin{tabular}[t]{@{}p{8.9cm}@{}}
$^{87}$Rb, five-level high-angular-momentum configuration:
$5S_{1/2}\!\rightarrow\!5P_{3/2}\!\rightarrow\!5D_{5/2}\!\rightarrow\!45F_{7/2}\!\rightarrow\!45G_{9/2}$.
\end{tabular}
\\

\rowcolor{LightIEEEblue}
3GPP Ambient IoT, n5/n8/n28-like operation
&
\begin{tabular}[t]{@{}l@{}}
n5 UL: 824--849 MHz\\
n8 UL: 880--915 MHz\\
n28 UL: 703--748 MHz\\
Bandwidth: 200/400/600/800 kHz
\end{tabular}
&
\begin{tabular}[t]{@{}p{8.9cm}@{}}
$^{87}$Rb, five-level high-angular-momentum configuration.
For n8, use
$5S_{1/2}\!\rightarrow\!5P_{3/2}\!\rightarrow\!5D_{5/2}\!\rightarrow\!45F_{7/2}\!\rightarrow\!45G_{9/2}$.
For n5 or n28, lower-frequency $nF_{7/2}\!\rightarrow\!nG_{9/2}$ transitions.
\end{tabular}
\\

Ambient IoT, 2.4-GHz-like operation
&
\begin{tabular}[t]{@{}l@{}}
Carrier: 2.4-GHz ISM\\
Bandwidth: 100 kHz--several MHz
\end{tabular}
&
\begin{tabular}[t]{@{}p{8.9cm}@{}}
$^{87}$Rb, four-level ladder configuration:
$5S_{1/2}\!\rightarrow\!5P_{3/2}\!\rightarrow\!95D_{5/2}\!\rightarrow\!96P_{3/2}$
\end{tabular}
\\

\rowcolor{LightIEEEblue}
Ambient/private IoT, upper-FR1 7-GHz-like operation
&
\begin{tabular}[t]{@{}l@{}}
Carrier: 6.95 GHz\\
Bandwidth: 150 kHz--several MHz
\end{tabular}
&
\begin{tabular}[t]{@{}p{8.9cm}@{}}
$^{133}$Cs, four-level ladder configuration:
$6S_{1/2}\!\rightarrow\!6P_{3/2}\!\rightarrow\!47D_{5/2}\!\rightarrow\!48P_{3/2}$
\end{tabular}
\\

\bottomrule
\end{tabular}

\vspace{0.5em}
\begin{minipage}{0.96\textwidth}
\footnotesize
\textit{Note:} The listed Rydberg states are representative candidates for system-level modeling. The exact transition frequencies should be refined using the quantum-defect model, Stark-map analysis, and experimental spectroscopy. For sub-GHz IoT links, a five-level high-angular-momentum configuration based on $nF_{7/2}\!\rightarrow\!nG_{9/2}$ is preferred, whereas conventional four-level ladder configurations are suitable for many GHz-range ambient/backscatter or private IoT reader links.
\end{minipage}
\end{table*}

\subsection{LPWA Networks}

Fig.\ref{NBIot}(a) shows a representative RAQR-empowered IoT network in which an infrastructure-side RAQR at the base station receives weak uplink signals from NB-IoT and LoRa devices. The atomic-level configuration adopted for each carrier band, summarized in Table~\ref{tab:iot_raq_level_selection}, follows directly from how Rydberg level spacings scale with the principal quantum number (n). For sub-6~GHz ambient IoT, a conventional four-level ladder \(nD \to (n+1)P\) with spacings \(\propto \frac{1}{n^3}\) is sufficient, as it lands naturally in the target band with \(n \approx 50\text{–}95\). Instead, for the 800–900~MHz LPWA bands, the same scheme would require \(n > 130\), which is theoretically feasible but experimentally untenable. To tackle this issue, we adopt a five-level high-angular-momentum \(^{87}\text{Rb}\) configuration, \(5S_{1/2}\rightarrow 5P_{3/2}\rightarrow5D_{5/2}\rightarrow nF_{7/2}\rightarrow nG_{9/2}\) with \(n \approx 45\).

As shown in Fig.~\ref{NBIot}(a), the IoT RF signal superimposed on a strong local oscillator perturbs the Rydberg atomic coherence, modulating both the absorption and phase of the probe beam. The modulated probe beam is then combined with a strong local optical beam, and balanced coherent optical detection (BCOD) recovers the AC signal transmitted by the IoT devices.

Under the parameter set listed in the caption of Fig. 2, the receiver is driven by a weak uplink field of \(E_s = 37\) µV/m. The RAQR output noise follows the model of \cite{GongRAQR}, which incorporates thermal, photon shot, and quantum projection noise. The NB-IoT signal is modeled as 12 fixed subcarriers with 15~kHz spacing, and the LoRa signal is modeled as a chirp-spread-spectrum waveform. Fig.~\ref{NBIot}(b) confirms that the RAQR-BCOD receiver recovers both LPWA waveforms successfully. Fig.~\ref{NBIot}(c) shows that, once the RF detuning is chosen to bias the receiver at the steepest slope of the atomic response, RAQR consistently exceeds the conventional RF receiver in effective SNR. The detuning, optical power, slope, and detection quadrature thus become the key design knobs of the atomic front end. 

\subsection{Ambient IoT Networks}
Fig.~\ref{NBIot}(a) also illustrates an ambient IoT architecture, where IoT devices communicate with an RAQR-enabled BS via harvested energy. This configuration is well-suited to passive and energy-constrained devices, which cannot support high transmit power. As summarized in Table~\ref{tab:iot_raq_level_selection}, ambient IoT may operate across both sub-GHz and sub-6 GHz bands. For sub-GHz operation, the five-level \(nF_{7/2}\!\rightarrow\!nG_{9/2}\) configuration is preferred, whereas the four-level ladder configurations suffice for the sub-6 GHz band.

Using the four-level configuration \cite{GongRAQR}, we evaluate whether RAQR-enabled BSs can support ambient IoT. 
Fig.~\ref{NBIot}(d) shows a MIMO SWIPT ambient IoT network in which an RAQR-enabled BS receives weak signals returned by passive or energy-constrained devices, benchmarked against a conventional RF network supporting IoT devices with 5~dBm transmit power. As shown, RAQR-enabled BS significantly enhances the weighted sum rate, but also extends the viable deployment range of ambient IoT devices, thereby providing a promising solution for energy-constrained networks. These results suggest that the receiver-side gain of RAQR is most pronounced precisely where ambient IoT is most constrained, i.e., low transmit power and long device-to-reader distances.
\section{Network-Level Perspective: RAQR in Cellular and Cell-Free IoT}

In this section, we evaluate the network-level coverage of RAQR-enabled cellular IoT networks and examine the cooperative reception of RAQR-empowered cell-free deployments. The Monte Carlo evaluation adopts the RAQR-BCOD setting of the previous section. The simulations use the \(^{133}{\rm Cs}\) configuration \(6S_{1/2}\!\rightarrow\!6P_{3/2}\!\rightarrow\!42D_{5/2}\!\rightarrow\!43P_{3/2}\). 
The equivalent cell radius $R_{\rm eq}=\sqrt{1/(\pi\lambda)}$ is about 178~m at $\lambda=10^{-5}$~m$^{-2}$ and 56~m at $\lambda=10^{-4}$~m$^{-2}$. 
Our central network-level finding is that the RAQR advantage is density-dependent: it is pronounced in sparse, noise-limited deployments where the receiver noise floor dominates, and it erodes as density rises and the stronger aggregate field drives the Rydberg medium into its nonlinear regime.

\begin{figure*}[t]
    \centering
    \includegraphics[width=\linewidth]{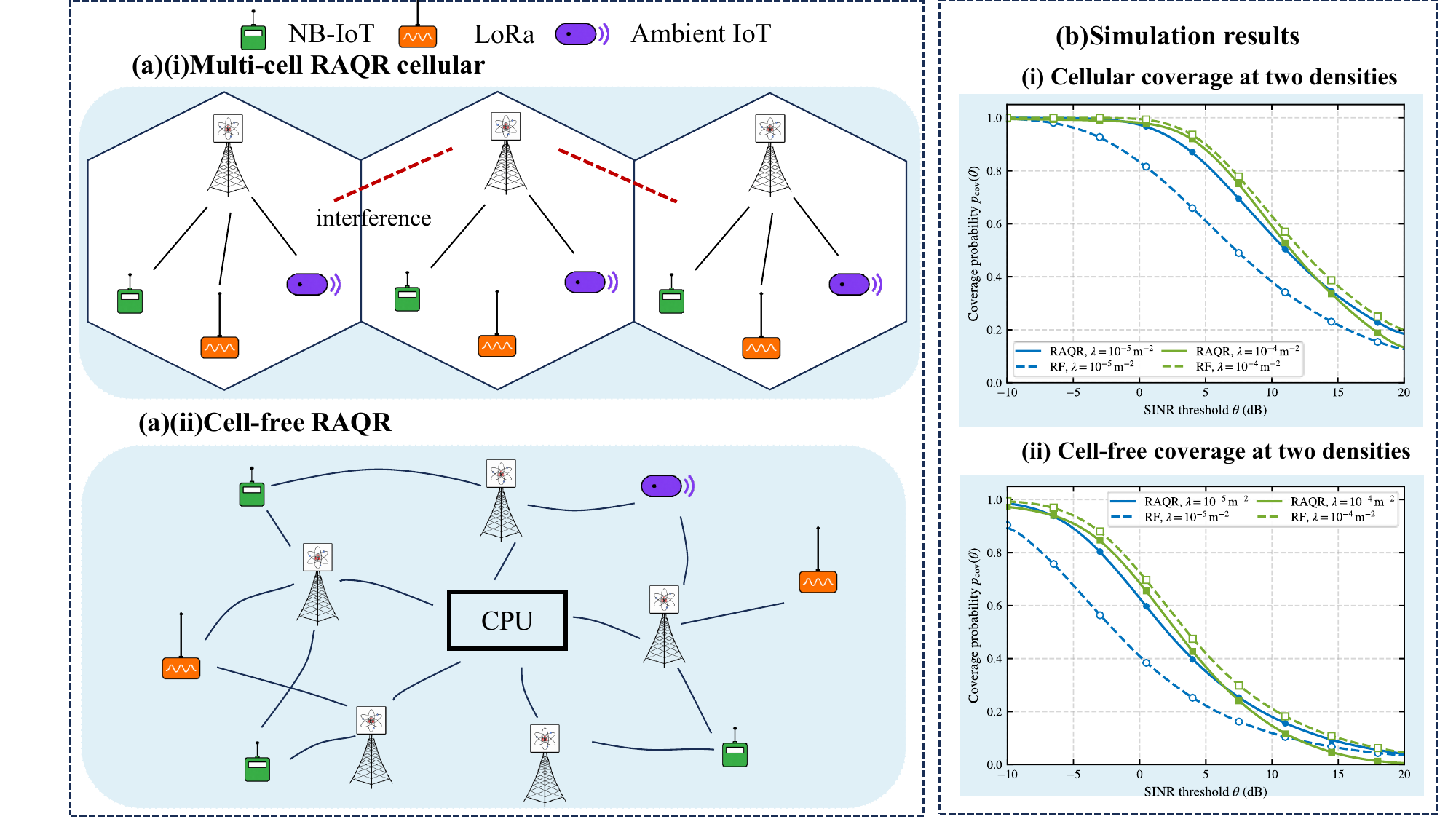}
    \caption{Network-level deployment of RAQR-empowered IoT. a) Typical multi-cell RAQR and cell-free RAQR; b) Coverage comparison with \(p_{\rm cov}(\theta) = \Pr(\mathrm{SINR} > \theta)\), \(\lambda_b = \lambda_u = \lambda\) denoting the common BS/AP and active-device density. Simulation parameters: four-level configuration with \(\Omega_p=2\pi\times3.35\)~MHz, \(\Omega_c=2\pi\times0.95\)~MHz, \(\Omega_{\rm LO}=2\pi\times1.22\)~MHz, and a \(30\)~mW local optical beam. RF baseline: NF = 5 dB over a \(B=800\)~kHz bandwidth on top of the thermal noise floor of (-174) dBm/Hz.}
    \label{fig:network}
\end{figure*}

\subsection{Cellular RAQR Networks}

As illustrated in Fig.~\ref{fig:network}(a)(i), the BSs form a Poisson point process of intensity $\lambda$, each with an RAQR-BCOD front end, and every device associates with its nearest BS. The vapor cells simultaneously serve NB-IoT, LoRa, and ambient terminals, and the field at each cell sums all co-band IoT uplinks.

Fig.~\ref{fig:network}(b)(i) reports the coverage probability at two densities. At $\lambda = 10^{-5}$~m$^{-2}$, the network is noise-limited, and the RAQR maintains half-coverage  (a coverage probability of $1/2$) at a threshold about 4 dB above the RF receiver, and thus the same coverage needs lower transmit power. At $\lambda = 10^{-4}$~m$^{-2}$, the two curves nearly overlap, with RF marginally better at moderate and high thresholds because the stronger aggregate field drives the Rydberg medium toward its nonlinear regime and saturates the RAQR SINR.

The RAQR retains a network-level noise advantage that narrows with device density. The practical takeaway is that RAQR's gain in cellular IoT is realized at the cell edge and in low-load regimes, where uplink noise dominates and aggregate-field nonlinearity is mild. Preserving this advantage at higher densities requires tuning the working point to the local field strength and exploiting narrowband selectivity to filter dense bands.

\subsection{Cell-Free RAQR Networks}

Fig.~\ref{fig:network}(a)(ii) shows a cell-free deployment where RAQR access points are distributed at intensity $\lambda$ and connected to a central processing unit through fronthaul. Each device is served by its $L=5$ nearest access points. At each access point, only the vapor cell, optical readout, and minimal digitization sit locally. The heavier baseband processing is aggregated at the CPU. NB-IoT, LoRa, and ambient terminals share the same access points.

Fig.~\ref{fig:network}(b)(ii) shows a similar trend under cell-free reception. At $\lambda = 10^{-5}$~m$^{-2}$, the RAQR keeps about a 4 dB half-coverage advantage, and at $\lambda = 10^{-4}$~m$^{-2}$ the RAQR and RF curves nearly overlap, with RF slightly better at moderate and high thresholds. The same narrowing in both topologies indicates that RAQR transduction, rather than the interference geometry, is the limiting factor.

In both architectures, the RAQR delivers the largest gain in sparse networks with weak uplinks. The cell-free deployment partially compensates for the density-induced erosion of this gain, since distributed vapor cells capture weak uplinks from spatially diverse locations while heavier baseband processing is offloaded to the CPU. The trade-off is that each access point experiences its own aggregate field and settles at its own working point, so joint calibration and nonlinear distortion compensation become essential. The broader implication is that cell-free architectures widen the operational envelope of RAQR-empowered IoT, but the atomic working-point management problem must be solved jointly with baseband design rather than relegated to the hardware layer.

\section{Open Challenges and Research Roadmap}

While the above sections demonstrate the substantial benefits of RAQR-empowered IoT networks, several open challenges remain on the path toward practical deployment. We outline the key directions below.

\begin{figure*}
       \centering
    \includegraphics[width=0.9\linewidth]{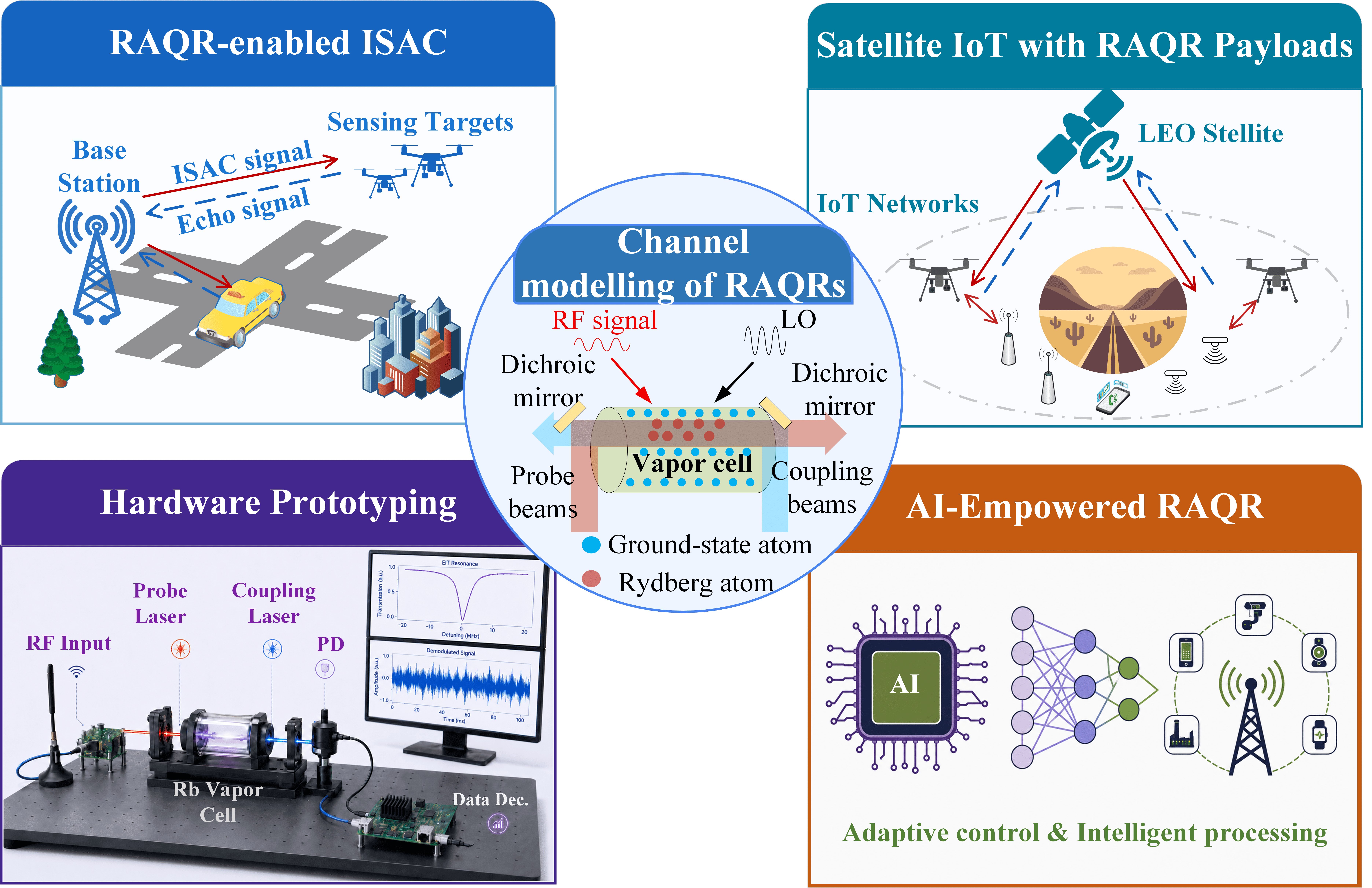}
    \caption{Open challenges and future research directions for RAQR-enabled IoT, covering channel modeling under non-Markovian regimes, RAQR-enabled ISAC, satellite IoT with onboard RAQR payloads, AI-empowered RAQR signal processing, and hardware prototyping.}
    \label{case}
\end{figure*}

\subsection{Channel Modeling of RAQRs}
Although recent work has begun to bridge the quantum-physics and communication communities, several gaps in RAQR channel modeling remain unresolved. First, different RAQR configurations, including the choice of atomic species, ladder structure, and operating point, yield markedly different equivalent channel models, and no unified modeling framework currently spans these design variants. Second, existing solutions to the Lindblad master equation typically invoke the rotating-wave approximation, adiabatic elimination, or perturbative expansions, each of which introduces systematic deviations between the modeled and the measured input–output response. Third, the Lindblad equation itself relies on the Born–Markov assumption and breaks down once the coupling-strength ratio leaves the weak-coupling regime.  In practical Rydberg vapor cells, dense atomic ensembles, structured electromagnetic environments, and strong RF excitation routinely push the effective coupling well beyond this limit. Closing these gaps will require a unified parameterized channel model that covers heterogeneous RAQR configurations, approximation-aware refinements of the Lindblad framework with quantified error bounds, and non-Markovian master-equation treatments capable of capturing bath memory effects in strong-coupling regimes \cite{trung}.

\subsection{RAQR-Enabled ISAC for IoT Networks}

Integrated sensing and communication (ISAC) is rapidly becoming a defining feature of next-generation IoT, and RAQRs are naturally suited to deliver this dual functionality. First, the RAQR response is governed by atomic transition frequencies and dipole moments, which are SI-traceable invariants and do not drift with temperature or component variation, eliminating the periodic recalibration that conventional RF sensors require. Second, its near-quantum-limited sensitivity recovers weak radar echoes from passive or low-cross-section targets that fall below the conventional radar noise floor, directly improving sensing accuracy. The two functions, however, place opposing demands on the atomic working point.  Communication operates in the small-signal regime, where the detector output scales with the local slope of the EIT-ATS transmission curve, and thus the steepest-slope bias maximizes the RF-to-optical conversion gain. On the contrary, sensing infers the absolute field from the Autler-Townes splitting, an SI-traceable mapping that holds only at the symmetric peak region. Promising avenues include adaptive biasing and multi-cell architecture design, wideband ISAC waveforms exploiting multi-level and multi-species atomic configurations to overcome the limited per-transition bandwidth, and cooperative atomic sensing across distributed vapor cells for high-resolution sensing.

\subsection{Satellite IoT with RAQR Payloads}
NTNs are a key enabler for global IoT coverage in remote, maritime, and aeronautical scenarios. However, their central bottleneck is the link budget caused by the severe free-space path loss and the constrained device-side transmit power, leaving even moderate-rate uplinks below the sensitivity of conventional RF payloads. RAQRs offer a direct response to this gap, since their sub-thermal noise floor pushes receiver sensitivity toward the standard quantum limit and can enable direct user-equipment access to satellites without extra power consumption at the user side \cite{11417150}. Nevertheless, deploying RAQRs onboard satellites raises several non-trivial challenges, including the space-grade robustness required of the vapor cell and laser stabilization subsystem, the dynamic detuning of atomic resonance induced by large LEO Doppler shifts, and the issues arising from massive concurrent IoT access, such as aggregate-field-induced nonlinearity, multi-user interference, and limited atomic dynamic range. Key open questions are space-resilient atomic front-end architectures, dynamics-aware atomic receiver design for satellite-ground links that jointly model large-scale Doppler effects, orbital dynamics, and atomic response, and multi-beam atomic signal processing together with cross-layer protocol design for massive-access NTN-IoT.

\subsection{AI-Empowered RAQR for IoT Networks}
Classical signal processing struggles to fully exploit the benefits of RAQRs, since the nonlinear EIT-ATS response, laser-phase noise, and atomic decoherence jointly yield a nonlinear and time-varying receiver model. Moreover, the atomic working point, RF detuning, local-oscillator power, and resource allocation across heterogeneous IoT devices also need  to be jointly tuned over a vast parameter space, which is intractable for conventional optimization approaches. In this context, AI plays a critical role in unlocking the full potential of RAQRs, owing to its modeling capacity and adaptivity in handling the inherently nonlinear, high-dimensional, and time-varying nature of atomic reception. Future research directions include physics-informed learning frameworks that embed the atomic response and quantum noise structure into model-driven detectors and decoders, AI-driven cross-layer optimization that jointly orchestrates atomic-level biasing, beamforming, and radio resource allocation, and reinforcement learning-based adaptive scheduling that tracks time-varying interference, device activity, and channel dynamics for massive IoT access.
 
\subsection{Hardware Prototyping}
Beyond modeling and algorithmic challenges, hardware integration remains the decisive bottleneck between current prototypes and deployable IoT infrastructure.  Current laboratory RAQR prototypes occupy optical benches with free-space lasers, bulky vapor cells, and discrete photodetectors, which are incompatible with the compact, low-cost form factor expected of IoT base stations and access points. Moreover, the laser stabilization subsystem dominates the power budget of present prototypes, while calibration and aging of the atomic medium further complicate long-term operation. Open questions are photonic integration of probe, coupling, and dressing lasers with micro-fabricated vapor cells toward shoebox-scale receivers, chip-scale frequency references and energy-efficient laser stabilization schemes for low-power operation, and self-calibration protocols leveraging the SI-traceable nature of atomic transitions to ensure long-term reliability.

\section{Conclusion}
RAQRs offer a physically distinct and complementary receiver-side path for next-generation IoT, combining near-quantum-limited sensitivity, immunity to the Chu limit, broad frequency tunability, and intrinsic narrowband selectivity. Our LPWA, ambient IoT, and stochastic-geometry case studies showed that this advantage is concrete yet conditional, with the largest gains arising in sparse, weak-uplink, energy-constrained regimes and narrowing once aggregate-field-induced nonlinearity sets in. Realizing the full promise of RAQR-empowered IoT will require co-design across channel modeling, ISAC, satellite payloads, AI-driven adaptation, and hardware integration. The atomic front end is unlikely to displace classical RF wholesale but is well positioned to become the complementary receiver of choice precisely where IoT is hardest, namely in weak-uplink, energy-starved, and infrastructure-sparse regimes.

\bibliographystyle{IEEEtran}
\bibliography{refs}

\end{document}